\documentstyle{article}
\begin{document}
\large
\baselineskip=24pt
\title{Accuracy of Discrete-Velocity BGK Models for the
Simulation of the Incompressible Navier-Stokes Equations\\[.2cm]}
\author{Marc B. Reider and James D. Sterling\thanks{
Permanent Address: Advanced Projects Research Incorporated,
5301 N.Commerce Ave., Suite A, Moorpark, CA 93021}\\
  [.2cm]{\em Center for Nonlinear Studies, MS-B258}\\
	{\em Los Alamos National Laboratory}\\
	{\em Los Alamos, NM 87545}\\[.2cm]}
\date{}
\maketitle
\begin{center}
\large{\bf ABSTRACT}
\end{center}
\baselineskip=12pt
Two discretizations of a 9-velocity Boltzmann equation with
a BGK collision operator are studied. A Chapman-Enskog expansion
of the PDE system predicts that the macroscopic behavior corresponds
to the incompressible Navier-Stokes equations with additional
terms of order Mach number squared.
We introduce a fourth-order scheme and compare results with those of
the commonly used lattice Boltzmann discretization and with
finite-difference schemes applied to the incompressible Navier-Stokes
equations in primitive-variable form. We numerically demonstrate
convergence of the BGK schemes to the incompressible
Navier-Stokes equations and quantify the errors associated
with compressibility and discretization effects.
When compressibility error is smaller than discretization error,
convergence in both grid spacing and time step
is shown to be second-order for the LB
method and is confirmed to be fourth-order for the fourth-order BGK solver.
However, when the compressibility error is simultaneously reduced as the
grid is refined, the LB method behaves as a first-order scheme in time.

\section{Introduction}
The Navier-Stokes equations of fluid mechanics may be derived as
the macroscopic behavior of hard-sphere particles with a Maxwellian
velocity distribution function whose evolution is
governed by Boltzmann's equation. However, vastly simpler
kinetic models can result in the same macroscopic behavior.
This idea led Frisch {\em et al} \cite{FHP} to the invention
of `minimalist' models for the purpose of numerical computation
of fluid flows. Pursuant to this
work, a significant number of papers have been published concerning
lattice gas automata (LGA) and the related lattice Boltzmann (LB)
method. As a subset of these we reference here two collections of
papers compiled by Doolen \cite{Dool1}\cite{Dool2}.

The LB method was proposed by McNamara and Zanetti
\cite{mcn} as a derivative of LGA's, but it may also be viewed as a
finite-difference method for the numerical simulation
of the discrete-velocity Boltzmann equation that makes use of
a BGK relaxation term instead of the full nonlinear collision
operator\cite{BGK}. Macroscopic Navier-Stokes
behavior is obtained by stipulating that the BGK relaxation
is towards an equilibrium velocity distribution function whose first
several velocity moments match those of a Maxwellian distribution.

Although the LB method has been used to simulate many physical phenomena,
its characterization as a numerical method has been primarily qualitative.
Several researchers have made quantitative comparisons of LB results with
traditional CFD methods \cite{benzi} \cite{jane} \cite{mart}
\cite{guy}. Comparisons of measured transport coefficients
have frequently been made (e.g. reference \cite{kad}) and
there has been some recent work concerning quantification of global error
and demonstration of convergence (e.g. reference \cite{skordos}).

In an effort to characterize the LB method and establish its validity
as a Navier-Stokes solver, we numerically demonstrate convergence
of a 9-velocity LB scheme to the incompressible Navier-Stokes equations.
The order of the method is determined and errors associated with
discretization and compressibility effects are identified.
We also introduce a finite-difference Boltzmann (FDB) scheme by
applying a fourth-order finite-difference method to the 9-velocity BGK
model and demonstrate convergence of this scheme
to Navier-Stokes behavior in the incompressible and continuum limits.
Comparisons are also made with 2nd and 4th order finite-difference methods
applied directly to the incompressible Navier-Stokes equations.

In section 2, we describe how the Chapman-Enskog expansion is used to
determine the macroscopic behavior associated with the discrete-velocity
BGK models. Error terms are discussed and the models are compared with
traditional computational fluid dynamics methods. In section 3, the
fourth-order finite difference Boltzmann scheme is introduced and its
differences from the traditional LB method are highlighted.
In Section 4 we present
the convergence studies and discuss the results. We conclude with some
comments regarding the overall accuracy, speed, and stability
of the discrete-velocity BGK models in comparison with other CFD methods.

\section{Discrete-Velocity BGK Macroscopic Behavior}
We begin by presenting a discrete-velocity BGK model
that uses nine velocities.
The particle populations, $f_{i}$, are associated with a zero-velocity
rest population, $f_{0}$, and with the
set of discrete velocities given by ${\bf
e}_{i}=c\{cos(\pi(i-1)/2),sin(\pi(i-1)/2)\}$ for $i=1,2,3,4$ and ${\bf
e}_{i}=\sqrt{2}c\{cos(\pi(i-\frac{9}{2})/2),sin(\pi(i-\frac{9}{2})/2)\}$
for $i=5,6,7,8$. The parameter $c$ is the characteristic particle speed
and is proportional to the speed of sound.
The mass and momentum of the fluid at a site are
obtained using the following sums over all nine velocities ($i=0,8$):

\noindent{Mass:}
\begin{equation}
n \equiv \sum_{i}f_{i}
\end{equation}
Momentum:
\begin{equation}
n{\bf u} \equiv \sum_{i}f_{i}{\bf e}_{i}.
\end{equation}

The BGK model presented here makes use of Boltzmann's equation with
the collision term replaced by a single-time relaxation towards an
equilibrium population denoted $f_{i}^{eq}$ \cite{BGK}.
The discrete-velocity model that describes the evolution of
$f_{i}$ is written
\begin{equation}
\frac{\partial f_{i}}{\partial t}
+{\bf e}_{i} \cdot {\bf \nabla}f_{i}= -\frac{1}{\tau} (f_{i}-f_{i}^{eq}).
\end{equation}
The value of $\tau$ in this model is inversely proportional to density
(assumed constant for incompressible flows) and
is assumed to be small so that for small nonequilibrium populations, the
right side of the above equation will be of $O(1)$. The
equilibrium populations of this model are given
in reference \cite{mart} as
\begin{equation}
f_{0}^{\rm eq} = \frac{4}{9}n-\frac{2}{3}n u^{2}
\end{equation}
\begin{equation}
f_{i}^{\rm eq} = \frac{1}{9}n+ \frac{n}{3 c^{2}}{\bf e}_{i}\cdot{\bf u}
+ \frac{n}{2 c^{4}}({\bf e}_{i}\cdot{\bf u})^{2}
-\frac{n}{6 c^{2}} u^{2}
\end{equation}
for $i=1,2,3,4$ and
\begin{equation}
f_{i}^{\rm eq} = \frac{1}{36}n+ \frac{n}{12 c^{2}}{\bf
e}_{i}\cdot{\bf u}
+ \frac{n}{8 c^{4}}({\bf e}_{i}\cdot{\bf u})^{2}
-\frac{n}{24 c^{2}} u^{2}
\end{equation}
for $i=5,6,7,8$.

A Chapman-Enskog procedure is applied to determine the macroscopic
behavior of this model. Details of this procedure are provided in
reference \cite{sterling}. The resulting continuity and momentum
equations follow.
\begin{equation}
\frac{\partial n }{\partial t} +
\frac{\partial n u_{\beta}}{\partial x_{\beta}} +
O(\varepsilon^{2})= 0
\end{equation}
\begin{equation}
n\frac{\partial u_{\alpha}}{\partial t}+nu_{\beta}\frac{\partial
u_{\alpha}}{\partial x_{\beta}}=
-\frac{\partial p}{\partial x_{\alpha}}+\frac{\partial}
{\partial x_{\beta}}(\mu(\frac{\partial u_{\beta}}
{\partial x_{\alpha}}+\frac{\partial u_{\alpha}}
{\partial x_{\beta}}))+O(\varepsilon^{2})+O(M^{3})
\end{equation}
where
\begin{equation}
\mu=\frac{\tau n c^{2}}{3}.
\end{equation}

Characteristic dimensionless parameters are the Mach number,
$M=\frac{\sqrt{3}U}{c}$ where $U$ is
a characteristic macroscopic flow speed,
the Knudsen number which is proportional to
$\varepsilon=\frac{c\tau}{L}$, where $L$ is a macroscopic flow
length, and the Reynolds number,
\begin{equation}
Re= \frac{n U L}{\mu} = \frac{\sqrt{3}ML}{\tau c}.
\end{equation}

We note that the Knudsen number is proportional to the Mach number so
that the $O(\varepsilon^{2})$ terms are of $O(M^{2})$.
Thus, equations (7) and (8) are the compressible
Navier-Stokes equations with additional terms that are of $O(M^{2})$.
Deviations from incompressible behavior are still present in the
compressible Navier-Stokes equations. These deviations are
associated with gradients of density and velocity field
divergence and are also of $O(M^{2})$ \cite{mart}\cite{majda}. We
group all three types of terms; Knudsen number squared, Mach
number cubed, and Navier-Stokes compressibility terms, into
`compressibility error'. As the Mach number approaches zero,
the above model should approach incompressible Navier-Stokes behavior.

Numerical simulations of the incompressible Navier-Stokes equations
may be performed by discretizing equation (3). We note that convergence
to the incompressible equations is obtained only if the compressibility
errors are smaller than the discretization error. Here we
discuss the commonly used LB discretization while in section 3 we
introduce a fourth-order discretization.

The lattice Boltzmann discretization of equation (3) consists of a
first order upwind discretization of the left side and the selection
of the lattice spacing $h$, and the time step $\Delta t$, to provide
an exact Lagrangian solution. This is accomplished by selecting
$\frac{h}{\Delta t}=c$ so that the discrete equation becomes
\begin{equation}
f_{i}({\bf x} + {\bf e}_{i}\Delta t,t + \Delta t)-f_{i}({\bf x},t)=
-\frac{\Delta t}{\tau}(f_{i}({\bf x},t)-f_{i}^{(0)}({\bf x},t)).
\end{equation}
While this appears to be a first-order method, it is actually second-order
if the second order terms in the truncation error
are considered to represent artificial viscosity.
More precisely, when a Taylor series expansion
is performed on the first term in the above equation,
all second order terms may be combined in a
manner that the value of $\tau$ in equation (9) may simply be replaced by
$\tau-\frac{\Delta t}{2}$ for the LB method.

The use of the second order discretization error to represent physics,
leaves only third-order terms as the truncation error so that
the method is effectively second order \cite{mario} \cite{sterling}.
Since the time step is proportional to the lattice spacing, the
scheme is second order in both time and space (again, if compressibility
effects are smaller than discretization error). These analytical
trends are numerically studied and verified in Section 4 below.

\section{Fourth Order Discrete Velocity BGK Model Solver}

In this section we present a fourth order finite difference
discretization of the discrete
velocity BGK equation.  In the previous section we described how
the Chapman-Enskog expansion can be used to recover the incompressible
Navier-Stokes equations from the BGK equation with a certain set
of discrete velocities.  We showed that if the
Mach number is sufficiently small, a numerical solver that
produces a higher order approximation to
the Boltzmann equation will also produce a higher order approximation
to the Navier-Stokes equations.

  We seek to discretize
\begin{equation}
  \frac{\partial f_{i}}{\partial t} + \vec{e}_{i} \cdot \nabla f_{i}
 =  -\frac{1}{\tau} (f_{i} - f^{eq}_{i})
\end{equation}
  We use centered differences to discretize the convective
terms.  This is done because of their simplicity to implement, and
because they have lower truncation errors than upwind schemes of equal
order.  The fourth order approximation  used for the convective term
can be written as
\begin{equation}
\frac{\partial f}{\partial x} \approx D^{0}_{x}f =
\frac{-f(x + 2h) + 8 f(x+h) -8f(x-h) + f(x-2h)}{12h}
\end{equation}
where $h$ is the mesh width.

  This produces a system of ordinary differential equations for the
distributions at each point :
\begin{equation}
\label{ode1}
  \frac{\partial f_{i}(x,y,t)}{\partial t} = -\vec{e}_{i} \cdot
(D^{0}_{x}f(x,y,t),D^{0}_{y}f(x,y,t))
  -\frac{1}{\tau} (f_{i}(x,y,t) - f^{eq}_{i}(x,y,t))
\end{equation}
  We discretize in time explicitly using the following
fourth order Runge-Kutta method.
\begin{eqnarray}
  \vec{k}^{0} & = &  \Delta t G(\vec{f}(t_{n}),t_{n}) \\
  \vec{k}^{1} & = &  \Delta t G(\vec{f}(t_{n}) + \frac{1}{2}\vec{k}^{0},
  t_{n} + \frac{1}{2}\Delta t) \\
  \vec{k}^{2} & = &  \Delta t G(\vec{f}(t_{n}) + \frac{1}{2}\vec{k}^{1},
  t_{n} + \frac{1}{2}\Delta t) \\
  \vec{k}^{3} & = &  \Delta t G(\vec{f}(t_{n}) + \vec{k}^{2},
  t_{n} + \Delta t) \\
  \vec{f}(t_{n+1}) & = &  \vec{f}(t_{n}) + \frac{1}{6}(\vec{k}^{0}
+ 2\vec{k}^{1} + 2\vec{k}^{2} + \vec{k}^{3})
\end{eqnarray}
where $G(\vec{x},t)$ is the right hand side of equation \ref{ode1}.
The use of centered differences on the convection term makes
some commonly used marching procedures such as Euler's and Heun's
method unstable in the absence of a dissipative term.  However both
third and fourth order Runge-Kutta methods have a region of stability
that contains a portion of the imaginary axis in the complex plane
and hence are stable for our discretization\cite{?}.  The larger stability
region of the fourth order Runge-Kutta method makes the computation more
efficient, and as a bonus we obtain fourth order accuracy in time.

This finite difference Boltzmann (FDB) scheme differs from the
traditional LB scheme.  The stream/collide process is replaced by the
combination of a finite
difference calculation of the convection terms and
a four step Runge-Kutta process for advancement in time.  A key difference
is the value of the viscosity.  Artificial viscosity produced by the
grid is not used to represent physics as in the LB method,
so the relationship between the relaxation parameter $\tau$
and the physical parameters can be
obtained directly from equation (10) as
\begin{equation}
  \tau = \frac{\sqrt{3}ML}{c Re}
\end{equation}
whereas for the LB method the relationship is
\begin{equation}
  \tau = \frac{\sqrt{3}ML}{c Re}+\frac{\Delta t}{2}
\end{equation}

We must examine the stability of this numerical scheme.  The value of
the time step $\Delta t$ is no longer set by the lattice size but is an
independent numerical parameter. In the
absence of a collision term the stability requirement for the
convective term is \cite{?}
\begin{equation}
  \Delta t < \frac{\sqrt{8} h}{e_{max}}
\end{equation}
where $e_{max}$ is the maximum absolute value of the discrete velocities.
However the collision term has an approximate  stability condition that
$\Delta t < \tau $, which is consistent with the earlier statement that
populations not be allowed to evolve far from equilibrium.
This condition can be quite restrictive, since for high Reynolds
numbers and low Mach numbers $\tau$ can be quite small.  One
does not encounter this problem with the LB method because,
as seen in equation (21),
$\Delta t$ is always of $O(\tau)$ for small Mach number.

\section{Test Problem and Results}

  We choose as our test problem the evolution of a decaying Taylor
vortex in a $2 \pi$ periodic domain.
The exact solution for the flow satisfies
\begin{eqnarray}
  u(x,y,t) & = & - \exp(-\nu t(w_{1}^{2} + w_{2}^{2}))
\cos(w_{1} x) \sin(w_{2} y) \\
  v(x,y,t) & = & \frac{w_{1}}{w_{2}}
  exp(-\nu t(w_{1}^{2} + w_{2}^{2})) \sin(w_{1} x) \cos(w_{2} y)
\end{eqnarray}
  Our tests were performed at $Re = 100$ with $w_{1} = 3$ and $w_{2} = 2$.

  For LB and FDB calculations, we must specify how to initialize the
populations $f_{i}$ from a given velocity field.  First, a pressure field
is generated from the velocity field by taking the divergence of the
momentum equation and solving for the pressure. Since the velocity field is
divergence free,
\begin{equation}
\nabla^{2}p=-\nabla\cdot(\bf u \cdot \nabla \bf u).
\end{equation}
We then use the equation of state to initialize the density to
$\rho=\rho_{0}+\frac{3p}{c^{2}}$. With the velocity and density
specified, we investigated two methods of initializing the
populations $f_{i}$. In the first, the populations are initialized
to the equilibrium distributions. In the second method, we add the
non-equilibrium populations to the equilibrium values using the
formula (see also \cite{skordos})
\begin{equation}
  f^{neq}_{i} = -\tau\Delta t(\frac{\partial f^{eq}_{i}}{\partial t} +
    \vec{e}_{i} \cdot \nabla f^{eq}_{i})
\end{equation}
where the derivatives are evaluated analytically using the known
exact solution for this flow.

\subsection{Rate of Convergence}

We first examine the spatial and temporal convergence rates for the
LB and FDB methods.  For a given Mach and Reynolds number,
the Chapman-Enskog expansion yields the compressibility
error discussed in Section 2 so that we are not exactly
calculating the solution to the incompressible Navier-Stokes equations.
However we expect that as we refine in time and space at a fixed Mach
number, our methods will be converging to some solution.  We first
examine the speed at which convergence to this solution occurs.

We do not have an exact solution for the discrete velocity Boltzmann
equation that is being simulated.  However we can calculate
convergence rates by looking at the difference in the values of
solutions computed on successive grids.  If we let
\begin{equation}
  E(h) = \| U(2h) - U(h) \|
\end{equation}
  where $U(h)$ is the solution calculated on a grid with mesh width $h$,
then the rate of convergence can be estimated by
\begin{equation}
  \rho = \log_{2} \frac{E(2h)}{E(h)}
\end{equation}

  In the remainder of this section we will use the $L^{2}$ norm to compute
our differences.  These differences will be evaluated at $t=1$.
  We first show in Table \ref{cmp1} the convergence rates for the LB
method  when
equilibrium populations are used for initialization.  Spatial and
temporal convergence are equivalent for this method because
the time step is a linear function
of the mesh size.  The Mach numbers presented in the tables are based on
the lattice speed instead of the sound speed and are therefore equal
to the actual Mach number divided by $\sqrt{3}$. We see that one
gets second order convergence, even for Mach numbers for which
compressibility effects are significant.

\begin{table}
\centering
\begin{tabular}{|c||c|c||c|c||c|c||} \hline
Lattice Size &  \multicolumn{2}{c||}{M = 0.25} &
\multicolumn{2}{c||}{M = 0.125} &\multicolumn{2}{c||}{M = 0.0625} \\
\cline{2-7}
             &  E(h)  &  $\rho$ & E(h) &  $\rho$ &
E(h) &  $\rho$ \\ \hline
$32 \times 32$ & .441 &      & .423 &      & .434  &  \\
$64 \times 64$ & .049 & 3.15 & .061 & 2.80 & .0629 & 2.79 \\
$128 \times 128$ & .010 &  2.37 & .015 & 1.98 & .0128 & 2.29 \\
$256 \times 256$ & .0024 & 2.03 & .0039 & 1.99 & .0031 & 2.05 \\   \hline
\end{tabular}
\caption{Convergence behavior for LB with first initialization method}
\vspace{0.25in}
\label{cmp1}
\end{table}

  One sees a different behavior when the second initialization
method is used.  This initialization appears to deteriorate convergence
behavior, and we see less than second order accuracy.  This failure to
reach second order decreases as Mach number decreases, so for very small
Mach number second order accuracy does result.  We note that a possible
cause of the loss of accuracy is that the artificial viscosity from the
grid that is used to attain second order accuracy is not taken into
account in this initialization method.

\begin{table}
\centering
\begin{tabular}{|c||c|c||c|c||c|c||} \hline
Lattice Size &  \multicolumn{2}{c||}{M = 0.25} &
\multicolumn{2}{c||}{M = 0.125} &\multicolumn{2}{c||}{M = 0.0625} \\
\cline{2-7}
             &  E(h)  &  $\rho$ & E(h) &  $\rho$ &
E(h) &  $\rho$ \\ \hline
$32 \times 32$ & .767 &      & .717 &    & .702 &  \\
$64 \times 64$ & .128 & 2.58 & .111 & 2.69 & .104 & 2.75 \\
$128 \times 128$ & .0339 & 1.92 & .0274 & 2.01 & .0238 & 2.13 \\
$256 \times 256$ & .0112 & 1.60 & .00787 & 1.80 & .00638 & 1.90\\   \hline
\end{tabular}
\caption{Convergence behavior for LB with second initialization method}
\vspace{0.25in}
\label{cmp2}
\end{table}

  Tables \ref{cmp3} and \ref{cmp4} show the space and time convergence
behavior for the FDB method.  We only present results at the
relatively high Mach number $M = .25$, but
one sees similar behavior throughout the entire range of Mach numbers.
The fourth order convergence behavior is quite clear, and small errors
are attained with much fewer lattice points than for the LB method.

\begin{table}
\centering
\begin{tabular}{|c|c|c|} \hline
Lattice Size & E(h)  &  $\rho$ \\ \hline
$32 \times 32$ & .899  &     \\
$64 \times 64$ & .0469 & 4.26 \\
$128 \times 128$ & .00249 & 4.24 \\ \hline
\end{tabular}
\caption{Convergence behavior in space for FDB}
\vspace{0.25in}
\label{cmp3}
\end{table}

\begin{table}
\centering
\begin{tabular}{|c|c|c|} \hline
Time Step & E(h)  &  $\rho$ \\ \hline
.0025 & $5.14 \times 10^{-5}$ &       \\
.00125 & $3.08 \times 10^{-6}$ & 4.06 \\
.000625 & $1.90 \times 10^{-7}$ & 4.02 \\ \hline
\end{tabular}
\caption{Convergence behavior in time for FDB}
\vspace{0.25in}
\label{cmp4}
\end{table}

\subsection{Effect of Initialization}

  It was expected that with increasing time,
the differences between the solutions
calculated with either initialization method would decrease.  However
we see that this is not the case.  As Table \ref{init} shows, for
a fixed grid size and Mach number, the difference
between the LB solutions computed with the two initialization methods
remains relatively constant for all time.   However this difference
decreases as the Mach number decreases and the mesh is refined.
It appears that in the limit of zero Mach number and infinite mesh
the two methods of initialization give equal solutions. But if one is
computing underresolved solutions, then it must be concluded that the
evolution of the computed velocity field is a function not just of the
initial velocity field but of the initial particle
distribution at each point.
Differences in initializations produce differences in the evolving
velocity field that do {\em not} decay in time.

\begin{table}
\centering
\begin{tabular}{|c|c|c|c|} \hline
Time & Grid Size &  Mach & Difference \\ \hline
.25 & $32 \times 32$ & 0.25 & .108 \\
1.00 & $32 \times 32$ & 0.25 & .116 \\ \hline
.25 & $64 \times 64$ & 0.25 & .0482 \\
1.00 & $64 \times 64$ & 0.25 & .0496 \\ \hline
.25 & $128 \times 128$ & 0.25 & .0268 \\
1.00 & $128 \times 128$ & 0.25 & .0271 \\ \hline
.25 & $256 \times 256$ & 0.25 & .0184 \\
1.00 & $256 \times 256$ & 0.25 & .0186 \\ \hline
.25 & $256 \times 256$ & 0.125 & .00632 \\
1.00 & $256 \times 256$ & 0.125 & .00489 \\ \hline
.25 & $256 \times 256$ & 0.0625 & .00250 \\
1.00 & $256 \times 256$ & 0.0625 & .00252 \\ \hline
\end{tabular}
\caption{Difference in solutions with different initialization methods}
\vspace{0.25in}
\label{init}
\end{table}

\subsection{Convergence to Incompressible Navier-Stokes Solutions}

  We now investigate the convergence of the LB and FDB methods to the
incompressible Navier-Stokes solutions to test problems.  For this
experiment we can now compare the numerical results to
the known exact solutions.  During simulation it
was found that the error oscillated in time.  In order to get a clear
picture of the accuracy of the methods, we computed the error at small
intervals up to $t=1$ and then averaged to compute an average error.
The average divergence was also computed to indicate the deviation from
incompressibility.

  Solutions were also calculated with a finite difference solver for
the Navier-Stokes equations.  Both a second and fourth order solver
that employed the primitive variables (velocity and pressure)
were used.  These results allow
us to compare the accuracy of the Boltzmann methods to methods that
discretize the Navier-Stokes equations directly.

  The results of this study are presented in figure 1.
The dotted lines represent the errors of the LB simulations as the mesh
is refined for fixed Mach numbers.  The solid lines are the FDB errors
for the same constant Mach numbers.  The dash-dotted lines represent
the errors in the finite difference solutions to the incompressible
Navier-Stokes equations.

  We observe that
convergence to the solution of the incompressible Navier-Stokes
equations is only attained by simultaneously refining both the mesh size
and decreasing the Mach number. As can be seen in figure 1,
the error for the LB method saturates at a certain level as the mesh is
refined.  These deviations from the exact solutions are caused by the
``compressibility errors'' that were discussed in section 2.
In table \ref{machcnv} we look at errors and divergences for
calculations on a $512 \times 512$ lattice at various Mach numbers.  At
this lattice size the discretization error is quite small, and
differences from incompressible Navier-Stokes behavior are due to
compressibility effects that are of $O(M^{2})$.
\begin{table}
\centering
\begin{tabular}{|c|c|c|} \hline
Mach Number & Velocity Error & Divergence \\ \hline
.25 & .0381 & .425 \\
.125 & .00935 & .102 \\
.0625 & .00202 & .0236 \\
.03125 & .000570 & .00542 \\ \hline
\end{tabular}
\caption{Error from Navier-Stokes as Mach Number varies for LB}
\vspace{0.25in}
\label{machcnv}
\end{table}

By simultaneously decreasing Mach number and mesh size, we can
demonstrate convergence to Navier-Stokes solutions. (see table
\ref{LBcnv}).
The convergence is second order in space, because halving the grid size
reduces the error by approximately a factor of four.
However, the time step is also halved
because it is proportional to the grid spacing, and the Mach number
is halved to keep compressibility error of the same size as the
discretization error.
The lower Mach number requires an increase in the time needed for the same
flow evolution to occur (eddy-turnover time). Thus, while the error is
reduced by a factor of four, the number of time steps
increases by a factor of four. Keeping compressibility error
equal to discretization error makes the
scheme effectively first-order in time.
\begin{table}
\centering
\begin{tabular}{|c|c|c|c|} \hline
Grid Size & Mach Number & Velocity Error &  $\rho$ \\ \hline
$16 \times 16$ & .25 & .875 &      \\
$32 \times 32$ & .125 & .129 & 2.76 \\
$64 \times 64$ & .0625 & .0293 & 2.13 \\
$128 \times 128$ & .03125 & .00719 & 2.03 \\
$256 \times 256$ & .015625 & .00178 & 2.02 \\ \hline
\end{tabular}
\caption{Error from Navier-Stokes as Mach Number varies for LB}
\vspace{0.25in}
\label{LBcnv}
\end{table}

  Figure 1 also shows the more rapid convergence of
the FLB method  to
Navier-Stokes solutions.  One can obtain errors near those of the LB
method with $\frac{1}{16}$ as many points.  This rapid convergence
leads to the dominance of compressibility errors at coarser grids than
for the LB method.

  Finally we present the errors from a second and fourth order finite
difference method for the incompressible Navier-Stokes equations
in both figure 1 and table
\ref{FDNS}.  One sees that the fourth order scheme clearly outperforms
all methods.  The second order scheme produces errors slightly smaller
than the LB method for an equal number of points.  For this method,
an elliptic equation must be solved at each time step.
However the time step of the calculation no longer is limited by
the Mach number so many fewer time steps are needed.

\begin{table}
\centering
\begin{tabular}{|c||c|c||c|c||} \hline
 Grid Size &  \multicolumn{2}{c||}{Second Order} &
 \multicolumn{2}{c||}{Fourth Order} \\ \cline{2-5}
              &  error  &  $\rho$ & error &  $\rho$ \\ \hline
 $16 \times 16$ & .275 &      & .0652 &       \\
 $32 \times 32$ & .0798 & 1.79 & .00464 & 3.81 \\
 $64 \times 64$ & .0208 & 1.94 & .000299 & 3.96 \\
 $128 \times 128$ & .00526 &  1.98 & .000019 & 3.99 \\ \hline
\end{tabular}
\caption{Convergence behavior for finite difference NS solvers}
\vspace{0.25in}
\label{FDNS}
\end{table}

\section{Conclusions}
Convergence of two different discretizations of a discrete-velocity
BGK model to the incompressible Navier-Stokes equations has been
numerically demonstrated. The model consists of PDEs
that describe the evolution of the velocity distribution function
associated with each discrete velocity. A Chapman-Enskog expansion
predicts that the macroscopic behavior of the model corresponds to the
incompressible Navier-Stokes equations with deviations of $O(M^{2})$
that are referred to as `compressibility error'.
Thus, if the Mach number is small enough, one should be able to
discretize the PDE using any preferred finite-difference method and
the convergence to incompressible Navier-Stokes behavior should occur
at the rate corresponding to the order of the finite-difference method.

We introduced a fourth-order finite-difference Boltzmann method (FDB) and
verified convergence to the exact solution of
the Navier-Stokes equations for
a decaying Taylor vortex flow. For small enough compressibility error,
fourth-order convergence was confirmed. Similarly, the commonly used
lattice Boltzmann discretization was confirmed to behave as a
second order scheme in both time and space when compressibility
error was smaller than discretization error and when the artificial
viscosity associated with second order discretization error was taken
to represent a physical viscosity.

The computational time of the LB and FDB methods is inversely proportional
to the Mach number. Therefore, it is most
efficient to choose the Mach number
so that compressibility error is equal to discretization error. As the
grid is refined, one must simultaneously decrease the Mach number as
illustrated in table 7. With this approach, the LB scheme behaves as a
first order scheme in time.

The stream/collide view of the LB scheme results in a naturally parallel
algorithm that is easy to program and allows the use of particle
reflection boundary conditions. However, with a small amount of
additional effort, one may implement a higher-order scheme that allows
coarser resolution for a given error or lower error for fixed resolution
in comparison with the LB method. Stability requirements
of the LB method are described in reference \cite{sterling}.
The stability requirements
of the FDB method require that the time step be on the
order of the BGK relaxation time. We hope to develop methods that have
a less stringent stability requirement while retaining higher order
accuracy. In addition, studies concerning the effect of boundary
conditions on accuracy are needed to complete the numerical
characterization of these methods.

\section*{Acknowledgments}
We thank F. J. Alexander, M. G. Ancona, D. O. Martinez, S. Chen,
and Q. Zou for helpful suggestions. The work was supported by
the U.S. Department of Energy at Los Alamos National Laboratory.
J.D.S. thanks G. Doolen, J. Rodgers, the Theoretical Division, and
the Center for Nonlinear Studies for sponsoring his visit to Los Alamos.

\section*{Figure Captions}
\noindent
Fig.1 : Error of Discrete-Velocity BGK Methods as function of Mach number
and grid resolution. Dashed lines represent the error of the LB method for
various Mach numbers. Solid lines represent the error of the FDB method. The
dashed-dotted lines are the errors of the finite-difference Navier-Stokes
solvers.

\end{document}